\documentclass[a4paper,12pt,onecolumn,notitlepage]{article}
\usepackage[pdftex,final]{graphicx}
\usepackage[pdfpagemode=None,bookmarksopen=false,colorlinks=true,linkcolor=blue,citecolor=blue]{hyperref}
\usepackage[T1]{fontenc}
\usepackage[all]{xy}
\usepackage{amsmath}
\usepackage{amssymb}
\usepackage{sectsty}
\usepackage{bibspacing}
\usepackage{rpkid}
\begin{document}
\thispagestyle{empty}
\begin{center}
\noindent
{\Large \textbf{The general form of $\gamma$-family\\of~quantum~relative~entropies}}\footnote{Published in: Open Systems and Information Dynamics \textbf{18}:2 (2011), 191-221.}\\
\ \\
{Ryszard Pawe{\l} Kostecki}\\
{\small\ \\
\textit{Institute of Theoretical Physics, University of Warsaw, Ho\.{z}a 69, 00-681 Warszawa, Poland}\\
\ \\
August 30, 2011}
\end{center}
\begin{abstract}{\small\noindent 
We use the Falcone--Takesaki non-commutative flow of weights and the resulting theory of non-commutative $L_p$ spaces in order to define the family of relative entropy functionals that naturally generalise the quantum relative entropies of Jen\v{c}ov\'{a}--Ojima and the classical relative entropies of Zhu--Rohwer, and belong to an intersection of families of Petz relative entropies with Bregman relative entropies. For the purpose of this task, we generalise the notion of Bregman entropy to the infinite-dimensional non-commutative case using the Legendre--Fenchel duality. In addition, we use the Falcone--Takesaki duality to extend the duality between coarse--grainings and Markov maps to the infinite-dimensional non-commutative case. Following the recent result of Amari for the Zhu--Rohwer entropies, we conjecture that the proposed family of relative entropies is uniquely characterised by the Markov monotonicity \textit{and} the Legendre--Fenchel duality. The role of these results in the foundations and applications of quantum information theory is discussed.
}\end{abstract}
\begin{center}{\small \textit{dedicated to Professor Roman S. Ingarden on the occasion of his 90$^{\mbox{\textit{th}}}$ birthday}}\end{center}
\section{Introduction\label{introduction.section}}
The information theory can be separated into two parts: information kinematics (described in terms of information geometry theory) and information dynamics (described in terms of inductive inference theory). Information kinematics incorporates the description of convex and smooth geometric structures on the spaces of information states, while information dynamics incorporates the methods of statistical inference such as Bayes' rule, maximum likelihood, and constrained relative entropic updating. Both parts of information theory share two underlying mathematical notions: information model and information deviation (relative entropy). Following the approach initiated by Ingarden et al \cite{IJKK:1982} and Eguchi \cite{Eguchi:1983,Eguchi:1985} (and continued in particular in \cite{Lesniewski:Ruskai:1999,Hasegawa:2003,Rodriguez:2003,Jencova:2004:entropies}), we think that the further structures of the information geometry theory, such as riemannian metrics or affine connections on information models, should follow from the additional requirements imposed on these two notions. However, this requires us to define information models and information deviations on the level of generality that covers all regimes of information geometry, ranging from the finite-dimensional commutative normalised case to the infinite-dimensional non-commutative non-normalised case. 

We define an \textit{information model} as a subset $\M$ of the space $\N_*^+$ of all positive, finite, normal, linear $\CC$-valued functions on a given commutative or non-commutative $W^*$-algebra $\N$.\footnote{An additional subscript \textvisiblespace$_1$ will be used to denote the subset of \textit{normalised} ($\omega(\II)=1$) elements, while an additional subscript \textvisiblespace$_0$ will be used to denote the subset of \textit{faithful} ($\omega(x^*x)=0\limp x=0\;\forall x\in\N$) elements. A function $\omega:\B\ra\KK$, where $\B$ is a Banach space and $\KK\in\{\RR,\CC\}$, is called \textit{normal} if{}f $\omega(\sup\F)=\sup_{x\in\F}\omega(x)$ for each directed filter $\F\subseteq\B$ with the upper bound $\sup\F$. The \textit{weak-$^*$ topology} on Banach dual $\B^\banach$ of $\B$ is defined as the weakest topology on $\B^\banach$ such that the linear functions $\B^\banach\ni\omega\mapsto\omega(x)\in\KK$ are continuous for every $x\in\B$. A $\KK$-linear function on $\B$ is normal if{}f it is continuous in weak-$^*$ topology. A $C^*$-algebra is a Banach space $\A$ over $\KK$ such that it is also a ring over $\KK$ with unit $\II$ and operation $^*$ defined by $(ab)^*:=b^*a^*$, $(\lambda_1a+\lambda b)^*:=\lambda_1^*a^*+\lambda_2^*b^*$ (where $^*:\KK\ra\KK$ denotes the complex conjugation), such that $\n{xy}\leq\n{x}\n{y}$, $\n{\II}=1$, $\n{x^*x}=\n{x}^2$ $\forall x,y\in\A$. A \textit{$W^*$-algebra} is a $C^*$-algebra $\N$ such that there exists a Banach space $\N_*$, called \textit{predual}, satisfying $\N_*^\banach=\N$. The set of normal linear $\KK$-valued (hence, weak-$^*$ continuous) functions on $\N$ is precisely a predual of $\N$. A \textit{weight} on a $W^*$-algebra $\N$ is a function $\omega:\N^+\ra[0,+\infty]$ such that $\omega(0)=0$, $\omega(x+y)=\omega(x)+\omega(y)$, $\lambda>0\limp\omega(\lambda x)=\lambda\omega(x)$. A weight $\omega$ is called \textit{semi-finite} if{}f $\forall x\in\N^+\;\exists y\in\N^+\;(x\geq y,\;y\neq0,\;\omega(y)<+\infty)$. A weight is called a \textit{trace} if{}f $\omega(u^*xu)=\omega(x)$ for all unitary $u\in\N$. The space of all semi-finite normal weights on $\N$ will be denoted $\W(\N)$.} Elements of $\N_*^+$ will be called \textit{states} (or \textit{information states}). If the additional normalisation condition is assmed, then the resulting space $\N_{*1}^+\subset\N_*^+$ consists of all normal algebraic states on $\N$. For commutative $\N$, $\N_*^+$ reduces to a space $L_1(\mho)$ of Daniell--Stone integrals on a Banach vector lattice of characteristic functions on Stone spectrum of a countably finite Dedekind complete maharanisable algebra $\mho$ that is canonically associated with $\N\iso L_\infty(\mho)$. In this case $\N_*^+$ reduces to a space $L_1(\mho)^+_1$ of all normal (hence, monotonically continuous) expectation functionals on $\N$, which bijectively corresponds to the space of all positive probability measures on $\mho$.

In the  commutative and normalised case $\N_*^+$ reduces to a space of all normal (hence, monotonically continuous) expectation functionals on $\N$, which is a space of Daniell--Stone integrals on a Banach vector lattice that bijectively corresponds to the space $\P$ of all positive probability measures on a countably finite Dedekind complete boolean algebra $\mho$ that is derived from $\N$. 

An \textit{information deviation} (also referred to as: negative relative entropy, information divergence, information gain, loss function, risk function, contrast functional) is defined as a map $D:\N_*^+\times\N_*^+\ra[0,\infty]$ such that $D(\omega,\phi)=0$ if{}f $\omega=\phi$. It plays a role of a non-symmetric distance functional on $\N_*^+$, and serves as a principal tool for quantification of relative information content of states.

The above definitions of information model and information deviation are suitable also for the purposes of information dynamics theory (see \cite{Kostecki:2011:AIP,Kostecki:2011:towards}). The main goal of information theory is to provide particular instances of inductive inference (information dynamics), so the derivation of the particular form of deviation functional from the requirements imposed on the information dynamics is an important task. This method of axiomatisation was advanced in particular in \cite{Shore:Johnson:1980,Shore:Johnson:1981,Johnson:Shore:1983,vanFraasen:1981,vanFraasen:Hughes:Harman:1986,Skilling:1988,Paris:Vencovska:1989,Paris:Vencovska:1990,Csiszar:1991,Uffink:1996,Caticha:2004}. However, we would like to keep the kinematic and dynamic properties of information theory separated for a while, allowing for a better understanding of information geometric kinematics on its own right. Hence, in this work we will consider only the information kinematics and, more specifically, the problem of unique selection of a particular form of information deviation functional by the mathematical conditions that are plausible from the perspective of information kinematic applications. There exist also many axiomatisations of information deviations provided this way (see e.g., \cite{Aczel:Daroczy:1975,Csiszar:1978,Donald:1986,Petz:1992:characterisation,Ohya:Petz:1993,Petz:1994:entropy,Csiszar:2008,Amari:2009:alpha:divergence}).

The virtue of (kinematic or dynamic) axiomatic characterisation is that it allows to equip the characterised object with an unambiguous justification that is grounded in the particular interpretation of the elementary axioms. Yet, a single mathematical axiomatisation can be equipped with arbitrary many conceptual interpretations, hence the relevance of this point can be questioned. Nevertheless, for the purposes of applications of the information theory, at least some minimal purely operational interpretation is required anyway, hence the axiomatic characterisation of the information geometric (and information dynamic) structures is important. The variety of different applications and axiomatisations suggests strongly that there is no `absolute' or `universal' method of model construction, as well as there is no `universal' choice of information deviation and no `universal' choice of a method of inductive inference. In consequence, the aim of mathematical inquiry is to identify various sets of conditions such that each set is necessary and sufficient to select some particular category of models and some particular information deviation (as well as some particular inductive inference rule). While the statistical inference (dynamics) and information geometric modelling (kinematics) based on different sets of conditions is incommensurable, each set on its own is a  valid universe for information theoretic (statistical, inferential, probabilistic, quantum) inquiry.

This paper aims at the problem of identification of a set of conditions underlying the particular information geometry theory that was originally developed along the lines of requirement of monotonicity of information geometric structures (deviations, metrics, connections) under the Markov morphisms of information models. The results of Chentsov \cite{Chentsov:1969,Chentsov:1972} showed that the constraint of monotonicity under Markov maps is strong enough to provide interesting characterisations of unique (Fisher--Rao \cite{Fisher:1925:theory:statistical:estimation,Rao:1945,Bhattacharyya:1943:b,Jeffreys:1946}) riemannian metrics and one-parameter family of (Chentsov--Amari \cite{Chentsov:1969,Chentsov:1972,Amari:1980}) affine connections in normalised finite-dimensional commutative case. Due to work of Morozova and Chentsov \cite{Morozova:Chentsov:1989} and final result of Petz \cite{Petz:1994:geometry,Petz:1996:monotone}, a family of riemannian metrics in normalised finite-dimensional non-commutative case was characterised by the same condition (an explicit integral representation of the Morozova--Chentsov--Petz family was given by Hansen \cite{Hansen:2006}). However, further results showed that monotonicity under Markov maps is insufficient for fine analysis of geometries of quantum states in non-commutative case, even in finite-dimensional normalised regime. This follows from the results of Jen\v{c}ov\'{a} \cite{Jencova:2001:geometry,Jencova:2003:flat,Jencova:2004:entropies} characterising the family of markovian monotone connections, as well as related results of Hasegawa and Petz \cite{Petz:Hasegawa:1996,Hasegawa:Petz:1997,Hasegawa:2003}, Grasselli and Streater \cite{Grasselli:Streater:2001:unique:AIP,Grasselli:Streater:2001:unique:IDAQ,Grasselli:2004} and Gibilisco and Isola \cite{Gibilisco:Isola:2001:characterisation:WY,Gibilisco:Isola:2003:WY,Gibilisco:Isola:2004} on characterisation of a subfamily of the Morozova--Chentsov--Petz metrics known as the Wigner--Yanase--Dyson metrics \cite{Wigner:Yanase:1963,Hasegawa:1993}, together with its boundary known as the Bogolyubov--Kubo--Mori metric \cite{Bogolyubov:1961,Kubo:1957,Mori:1956}. This leads to a question about the additional principle of structural properties which would enable an improved characterisation.

As already remarked, Ingarden et al \cite{IJKK:1982} and Eguchi \cite{Eguchi:1983,Eguchi:1985} observed that the structures of riemannian metric and affine connections on information manifold can be derived from information deviation functional by suitable differentiation of the arguments and passing to the limit with one argument converging to other. In particular, the Fisher--Rao metric and the family of Chentsov--Amari connections can be derived in finite-dimensional commutative case by differentiation of the family $D_p$ of Zhu--Rohwer deviations \cite{Zhu:Rohwer:1995,Zhu:Rohwer:1997,Zhu:Rohwer:1998}. This approach was continued also in the finite-dimensional non-commutative case \cite{Lesniewski:Ruskai:1999,Hasegawa:2003,Jencova:2004:entropies}. This leads us to wonder whether (and which of) the properties of information deviation could be used to characterise the structures of information geometry. Amari \cite{Amari:2009:alpha:divergence} recently showed that the Zhu--Rohwer deviations are characterised (under some mild auxiliary conditions) as an intersection of the families of Csisz\'{a}r--Morimoto \cite{Csiszar:1963,Csiszar:1967:fdiv,Morimoto:1963} and Bregman \cite{Bregman:1967} deviations. While the former family is characterised \cite{Csiszar:1978,Csiszar:1991} by monotonicity under Markov maps, the latter is characterised \cite{Jones:Byrne:1990,Csiszar:1991} by additive decomposition under projection on the subset that is `orthogonal' to the projection in the sense of Legendre duality.

This leads us to consider duality as a principle of characterisation that is equally fundamental as markovian monotonicity. However, as opposed to perspective present in standard treatments such as \cite{Nagaoka:Amari:1982,Eguchi:1983,Eguchi:1985,Lauritzen:1987:statistical:manifolds,Lauritzen:1987:conjugate:connections,Amari:Nagaoka:1993,Zhang:Matsuzoe:2009}, we require also that the dualistic properties should be independent of differential properties, because the former are essentially related to variational component of the theory, which in infinite-dimensional setting becomes quite independent of the smooth component of the theory. Consideration of non-smooth infinite-dimensional duality as a principle that is as important as Markov monotonicity leads us to deny the foundational role of differential geometric Norden--Sen duality \cite{Norden:1945,Sen:1944} in favour of variational convex Legendre--Fenchel duality \cite{Fenchel:1949,Broensted:1964}. Turning this idea into structure, we define the generalised Bregman deviation in terms of the convex Legendre--Fenchel duality on the dualised vector spaces, with no differentiability and no topological continuity properties required a priori. 

Jen\v{c}ov\'{a} \cite{Jencova:2003:arXiv,Jencova:2005} and Ojima \cite{Ojima:2004} introduced a family $D_p$ of quantum deviation functionals on $\N_*^+$, defined using the dual structure of non-commutative Araki--Masuda $L_p(\N,\phi)$ spaces \cite{Araki:Masuda:1982,Masuda:1983}. These deviations are well-defined in the non-commutative infinite-dimensional case, and belong to an intersection of the families of generalised Bregman deviations and Markov monotone Petz deviations \cite{Petz:1985:quasientropies,Petz:1986:quasi:finite} (the latter are non-commutative generalisation of the Csisz\'{a}r--Morimoto deviations). In particular cases, the Jen\v{c}ov\'{a}--Ojima deviations reduce to relative entropies of Hasegawa \cite{Hasegawa:1993} and Umegaki--Araki \cite{Umegaki:1962,Araki:1976:relative:entropy:I,Araki:1977:relative:entropy:II}. However, the Araki--Masuda $L_p(\N,\phi)$ space \textit{is not} a canonical non-commutative $L_p$ space construction, due to its dependence on an arbitrary weight $\phi$. In consequence, the Jen\v{c}ov\'{a}--Ojima deviation \textit{is not} a canonical non-commutative generalisation of the Zhu--Rohwer deviation, because the construction of the former depends on the choice of reference weight $\phi$, while the latter is independent of any choice of this kind.

We propose a solution to this problem, which is based on the Falcone--Takesaki theory \cite{Falcone:Takesaki:2001} of non-commutative flow of weights and associated construction of non-commutative $L_p(\N)$ spaces that does not depend on a choice of reference weight. Using its remarkable properties, we define the family $D_p$ of information deviation (relative entropy) functionals which is the canonical generalisation of Jen\v{c}ov\'{a}--Ojima \textit{and} Zhu--Rohwer families of deviations. We show that our family $D_p$ of information deviations belongs not only to Petz's class of Markov monotone deviations, but also to the class of generalised Bregman deviations. As a non-commutative counterpart to the result of Amari \cite{Amari:2009:alpha:divergence}, we conjecture that the family of $D_p$ deviations is uniquely characterised as intersection of Markov monotone deviations with generalised Bregman deviations. This requirement states that quantum information deviations should be monotone under (Markov) coarse-graining of information \textit{and} should allow additive decompositions under (Legendre--Fenchel) orthogonal projections on closed convex affine information subspaces.

This way the family $D_p$ of information deviations on $\M\subseteq\N_*^+$ becomes equipped with an information theoretic interpretation, and the same is true for further information geometric structures on $\M$ derived from these deviations, such as riemannian metrics and affine connections. This interpretation refers to the notion of information as a fundamental entity, and is not involved in the conceptual problems associated with the interpretations of probability and the interpretations of Hilbert space based quantum theory.

From the perspective of applications, this allows to discuss the algebraic models of quantum statistical mechanics and quantum field theory as particular cases of quantum information theory. The use of the Falcone--Takesaki theory and the Legendre--Fenchel duality advocated in this paper is a key novel technique that allows to derive various new results in the infinite-dimensional non-commutative case, where neither the suitable differentiability properties nor the description in terms of density operators is available. From the perspective of foundations, this allows to provide a major change of perspective: instead of separate consideration of statistical theory and quantum theory, with their separate problems of model construction, one obtains a single framework covering all theories at once. The model construction techniques based on spectral theory can be replaced by the model construction techniques based on quantum information geometry (which includes the spectral theory as a special case), with the  family $D_p$ of information deviations playing the central role (and replacing the role played by the Hilbert space norm). 
\section{Information models}
\subsection{Non-commutative flow of weights\label{FT.section}}
In what follows, we will often identify the $W^*$-algebra $\N$ with its standard representation von Neumann algebra $\pi(\N)$ on the standard representation Hilbert space $\H$ \cite{Haagerup:1975:standard:form}. Every $W^*$-algebra $\N$ has a unique standard representation, up to unitary isomorphism. This representation is faithful ($\ker(\pi)=\{0\}$) and is unitarily isomorphic with the Gel'fand--Na\u{\i}mark--Segal representation \cite{Segal:1947:irreducible} for a pair $(\N,\omega)$, whenever $\omega\in\N_{*0}^+$ or $\omega\in\W_0(\N)$. While the set $\N_{*0}^+$ is non-empty if{}f $\N$ is countably finite (i.e., it is isomorphic to a von Neumann algebra possessing a cyclic and separating vector), $\W_0(\N)\neq\emptyset$ for every $W^*$-algebra $\N$.

Two crucial elements of the Falcone--Takesaki theory \cite{Falcone:Takesaki:2001} are the \textit{core} von Neumann algebra $\core$ associated functorially to any von Neumann algebra $\N$, and the resulting canonical construction of non-commutative $L_p(\N)$ spaces, which is independent of any reference weight or state.

For $x,x'\in\N$, $\omega,\omega'\in\W_0(\N)$, and $\left(\frac{\DD\omega}{\DD\omega'}\right)_t$ denoting Connes' cocycle \cite{Connes:1973:classification} (which is a generalisation of the Radon--Nikod\'{y}m derivative to the non-commutative case), one defines the equivalence relation $\sim_t$ on $\N\times\W_0(\N)$ by
\begin{equation}
        (x,\omega)\sim_t(x',\omega')\iff x'=x\left(\frac{\DD\omega}{\DD\omega'}\right)_t.
\end{equation}
The equivalence class $(\N\times\W_0(\N))/\sim_t$ is denoted by $\N(t)$, and its elements are denoted by $x\omega^{it}$. The definitions $(x\omega^{it}+y\omega^{it}):=(x+y)\omega^{it}$, $\lambda(x\omega^{it}):=(\lambda x)\omega^{it}$ for $\lambda\in\CC$, and $\n{x\omega^{it}}:=\n{x}$ equip $\N(t)$ with the structure of the Banach space, which is isometrically isomorphic to the Banach space structure of $\N$. By definition, $\N(0)$ is isomorphic to $\N$ also as a von Neumann algebra. The product topology on $\N\times\RR$ (with $\N$ endowed with the weak-$^*$ topology or Arens--Mackey topology, but not with norm topology) allows to use the bijections $\N(t)\ni x\varphi^{it}\mapsto(x,t)\in\N\times\RR$ to form Fell's Banach $^*$-algebra bundle $F(\N):=\coprod_{t\in\RR}\N(t)$ over $\N\times\RR$ \cite{Fell:1969,Fell:1977}. The \textit{core} von Neumann algebra $\core$ is constructed from $\N$ as a result of action of the Banach $^*$-algebra of suitable sections of $F(\N)$ on a suitably defined auxiliary Hilbert space (see \cite{Falcone:Takesaki:2001} for details). The construction of $\core$ does not depend on any weight $\W_0(\N)$. However, for every choice of a particular $\omega$, there exists a unitary map providing the isomorphism $\core\iso\N\rtimes_{\sigma^\omega}\RR$, where $\sigma^\omega(\cdot)=\Delta_\omega^{it}(\cdot)\Delta_\omega^{-it}$ denotes the Tomita--Takesaki modular automorphism of $\N$ associated with $\omega$ \cite{Tomita:1967:a,Takesaki:1970}, $\N\rtimes_{\sigma^\omega}\RR$ is a crossed product von Neumann algebra acting on $\H\otimes L_2(\RR,dt)$ \cite{Takesaki:1973:duality,vanDaele:1978}, and $\H$ is a standard representation Hilbert space of $\N$ \cite{Haagerup:1975:standard:form}. 

The one-parameter automorphism group of $F(\N)$, defined by
\begin{equation}
        \tilde{\sigma}_s(x\phi^{it}):=e^{-its}x\phi^{it}\;\;\;\;\forall x\phi^{it}\in\N(t),
\label{tilde.sigma.def}
\end{equation}
extends uniquely to an automorphism group $\tilde{\sigma}_s:\core\ra\core$. It allows to define a \textit{grade} $\grad(T)$ of a closed densely defined operator $T$ affiliated with $\core$ as such $\gamma\in\CC$ that
\begin{equation}
        \tilde{\sigma}_s(T)=e^{-\gamma s}T\;\;\;\;\forall s\in\RR.
\label{grade.def}
\end{equation}
If $\grad(T)=0$, then $T$ is bounded, but if $\re\grad(T)\neq0$, then $T$ is unbounded. The action of $\tilde{\sigma}_s$ on $\core$ is integrable over $s\in\RR$, and $I_{\tilde{\sigma}}(x):=\int_\RR ds\tilde{\sigma}_s(x)$, $x\in\core^+$, is an operator valued weight from $\core$ to $\N$ (for details on operator valued weights, see \cite{Haagerup:1979:ovw1,Haagerup:1979:ovw2,Falcone:Takesaki:1999}). This allows to equip $\core$ with a faithful normal semi-finite trace $\taucore$, defined by
\begin{equation}
        \taucore_\varphi(x):=\lim_{\epsilon\ra+0}\varphi\circ I_{\tilde{\sigma}}(\varphi^{-1/2}(1+\epsilon\varphi^{-1})^{-1/2}x\varphi^{-1/2}(1+\epsilon\varphi^{-1})^{-1/2}).
\label{canonical.trace}
\end{equation}
This definition is independent of the choice of weight ($\taucore_\varphi=\taucore_\psi\;\forall\varphi,\psi\in\W_0(\N)$), which allows to write $\taucore$ instead of $\taucore_\varphi$. Moreover, $\taucore$ has the scaling property
\begin{equation}
        \taucore\circ\tilde{\sigma}_s=e^{-s}\taucore\;\;\;\;\forall s\in\RR.
\label{scaling}
\end{equation}
This allows to consider $\taucore$ as a \textit{canonical trace} of $\core$. Falcone and Takesaki call the system $(\core,\RR,\tilde{\sigma},\taucore)$ the \textit{non-commutative flow of weights}. 
\subsection{Non-commutative $L_p(\N)$ spaces}
Let $\MMM^p(\core)$ denote the space of all $\taucore$-measurable operators of grade $1/p$ affiliated to $\core$ (for details on measurable operators, see \cite{Segal:1953,Nelson:1974,Terp:1981}), and let the set $\mmm^+_{I_{\tilde{\sigma}}}\subseteq\core^+$ be defined as
\begin{equation}
        \mmm^+_{I_{\tilde{\sigma}}}:=\{x^*y\in\core\mid\;\n{I_{\tilde{\sigma}}(x^*x)}<\infty,\;\n{I_{\tilde{\sigma}}(y^*y)}<\infty\}.
\end{equation}
Given weight $\varphi\in\W(\N)$ and a canonical trace $\taucore$, the construction
\begin{equation}
        h^{it}_\varphi:=\left(\frac{\DD(\varphi\circ I_{\tilde{\sigma}})}{\DD\taucore}\right)_t\;\;\forall t\in\RR
\label{haagerup.corresp}
\end{equation}
defines the operator valued maps $\varphi\mapsto h_\varphi$ on $\W(\N)$, called the Haagerup correspondence \cite{Haagerup:1979:ncLp,Yamagami:1992}. From the Pedersen--Takesaki non-commutative generalisation \cite{Pedersen:Takesaki:1973} of the Radon--Nikod\'{y}m theorem it follows that $h_\varphi$ is a unique positive self-adjoint operator affiliated with $\core$ which satisfies
\begin{equation}
        \varphi\circ I_{\tilde{\sigma}}(\cdot)=\taucore(h_\varphi\;\cdot).
\end{equation}
Moreover, for any closed and densely defined positive operator $T$ affiliated with $\core$ with $p:=\re\grad(T)>0$ there exists a unique weight $\varphi$ such that $h_\varphi$ is of grade $1$ and $h_\varphi=|T|^{1/p}$. Given a polar decomposition of $\varphi=u|\varphi|$ as a linear form, the weight $\varphi$ is finite ($|\varphi|(\II)<\infty$) if{}f $h_\varphi$ is $\taucore$-measurable \cite{Nelson:1974,Haagerup:1979:ncLp}. Hence, $h_\varphi$ can be considered as (a reference-independent) `operator density' of $\varphi$. For $\varphi\in\N_*^+$, an assignment $T(\varphi):=uT(|\varphi|)$ defines a unique extension of the Haagerup correspondence map to a natural bijection (linear isomorphism)
\begin{equation}
        \N_*\ni\omega\mapsto T(\omega)\in\MMM^1(\core).
\label{SDHFT.iso}
\end{equation}
Falcone and Takesaki show that the integral on $T$, given by
\begin{equation}
        \int T:=\taucore(a^{1/2}Ta^{1/2}),\;\;T\in\MMM^1(\core),\;\;a\in\mmm^+_{I_{\tilde{\sigma}}},\;\;\;I_{\tilde{\sigma}}(a)=1,
\end{equation}
is well defined. While $\taucore$ takes only the $+\infty$ value on non-zero elements of $\MMM^1(\core)$, the integral $\int$ takes finite values. This allows to extend (\ref{SDHFT.iso}) to an isometric isomorphism of Banach spaces, with the norm on $\MMM^1(\core)$ defined by $\n{T}_1:=\int|T|$, and with $\n{T(\varphi)}_1=\varphi(\II)$. The duality pairing between Banach spaces $\N$ and $\MMM^1(\core)$ that identifies $\N_*$ with $\MMM^1(\core)$ is given by the bilinear form
\begin{equation}
        \N\times\MMM^1(\core)\ni(x,T)\mapsto\duality{x,T}_{\core}:=\int xT\in\CC.
\label{int.duality.one}
\end{equation}
The non-commutative $L_p(\N)$ spaces for $p\in[1,\infty[$ are defined as the spaces $\MMM^p(\core)$ equipped with, and Cauchy complete in, the norm
\begin{equation}
        \n{T}_p:=\left(\int|T|^p\right)^{1/p},\;\;T\in\MMM^p(\core).
\end{equation}
By (\ref{SDHFT.iso}) and (\ref{int.duality.one}), $L_1(\N)\iso\N_*$, and it is natural to define $L_\infty(\N)\iso\N\iso\N(0)$, using the definition (\ref{grade.def}) of grade with $\tilde{\sigma}_s(T)=T$ for $\grad(T)=0$. The space $L_2(\N)$ is isometrically isomorphic to the Hilbert space $\H$ of standard representation of $\N$ \cite{Haagerup:1975:standard:form}, and the inner product on $L_2(\N)$ defined by
\begin{equation}
L_2(\N)\times L_2(\N)\ni(T_1,T_2)\mapsto\s{T_1,T_2}_{L_2(\N)}:=\int T_2^*T_1\in\CC
\label{FT.Hilbert.space}
\end{equation}
allows to identify them. For any choice of $\omega\in\W_0(\N)$ (or $\omega\in\N_{*0}^+$), $L_2(\N)$ is unitarily isomorphic to the GNS Hilbert space $\H_\omega$. By definition, all $L_p(\N)$ for $p\in[1,\infty]$ are Banach spaces. For any choice of a reference weight $\psi\in\W_0(\N)$, they are isometrically isomorphic to the non-commutative $L_p(\N,\psi)$ spaces of Kosaki and Terp \cite{Kosaki:1984:ncLp,Terp:1982}, Araki and Masuda \cite{Araki:Masuda:1982,Masuda:1983}, Connes and Hilsum \cite{Connes:1980,Hilsum:1981}, and Haagerup \cite{Haagerup:1979:ncLp,Terp:1981}. If the von Neumann algebra $\N$ is semi-finite and some faithful normal semi-finite trace $\tau_\psi$ on $\N$ is chosen, then the Falcone--Takesaki $L_p(\N)$ spaces are isometrically isomorphic and order-isomorphic to non-commutative $L_p(\N,\tau_\psi)$ spaces developed in \cite{Dye:1952,Segal:1953,Dixmier:1953,Ogasawara:Yoshinaga:1955,Kunze:1958,Stinespring:1959,Nelson:1974,Yeadon:1975}.

The duality (\ref{int.duality.one}) extends to non-commutative $L_p(\N)$ space duality, given by the bilinear map
\begin{equation}
        L_p(\N)\times L_q(\N)\ni(S,T)\mapsto\duality{S,T}_{\core}:=\int ST\in\CC,
\label{dual.pairing.Lp.FT}
\end{equation}
with $1/p+1/q=1$, where $p\in\{z\in\CC\mid\re(z)>0\}$, and $L_{1/it}(\N)=\N(t)$ for all $t\in\RR$. This way the Falcone--Takesaki theory incorporates also Izumi's \cite{Izumi:1996,Izumi:1997,Izumi:1998} complex extension of the weight-dependent Kosaki--Terp theory.

The spaces $\MMM^p(\core)$, $p\in\CC\setminus\{0\}$, embed naturally into the $*$-algebra $\MMM(\core)$ of all $\taucore$-measurable operators affiliated to $\core$. Due to the properties of grade function, the elements of $\MMM(\core)$ possess remarkable algebraic properties. The grade function satisfies:
\begin{equation}
        \begin{array}{c}
        \grad(T^*)=(\grad(T))^*,\;\;\;\grad(\overline{ST})=\grad(S)+\grad(T),\\
\grad(|T|)=\re(\grad(T))=\frac{1}{2}(\grad(T)+\grad(T)^*),\\
\re(p)<0\limp L_{\frac{1}{p}}(\N)=\{0\},\;\;\;\re(\grad(T))\geq0\limp|T|^{\frac{1}{\re(\grad(T))}}\in\N^+_*,
        \end{array}
\end{equation}
where $\overline{ST}$ is the closure of $ST$, and the last property is understood via the bijection between $\omega\in\N_*^+$ and the corresponding $h_\omega\in L_1(\N)$. If  $\phi,\psi\in\N_*^+$ and $x,y\in\N$, then the elements $x\phi^\gamma$ and $y\psi^\lambda$ can be added and multiplied freely inside $\MMM(\core)$ for all $\gamma,\lambda\in\CC$ such that $\re(\gamma)\geq0$ and $\re(\lambda)\geq0$. Moreover, for any $\{T_i\}_{i=1}^n\subseteq\MMM(\core)$ such that $\sum_i\grad(T_i)=1$,
\begin{equation}
        \prod_iT_i\in L_1(\N),\;\;\int T_1\cdots T_n=\int T_nT_1\cdots T_{n-1},\;\;\n{T_1\cdots T_n}\leq\n{T_1}\cdots\n{T_n},
\end{equation}
holds. In particular, one can use the formal algebraic relations of $\N(t)$ to rewrite the Tomita--Takesaki modular automorphism as $\sigma_{-t}^\phi(x)=\Delta^{-it}_\phi x\Delta^{it}_\phi=\phi^{-it}x\phi^{it}$, and to rewrite Connes' cocycle as  $\left(\frac{\DD\omega}{\DD\phi}\right)_t=\Delta_{\omega,\phi}^{it}\Delta_{\phi}^{-it}=\omega^{it}\phi^{-it}$, where $\Delta_{\omega,\phi}$ denotes the Connes--Digernes relative modular operator \cite{Araki:1973:relative:hamiltonian,Connes:1974,Digernes:1975}. These remarkable algebraic properties were first observed by Woronowicz \cite{Woronowicz:1979} and were later analysed in details by Yamagami \cite{Yamagami:1992}. 
\subsection{Comparison with traditional approaches}
With these results at hand, let us provide a brief discussion of the definition of information model which was given in Introduction.

From the mathematical point of view, the setup of $W^*$-algebras and their preduals is uniquely selected by three requirements:
\begin{enumerate}
\item[1)] the complex non-commutative generalisation of the abstract (Daniell--Stone) integration theory on abstract (Riesz) vector lattices should be provided in terms of the abstract $*$-algebras,
\item[2)] the duality pairing between the $*$-algebra and the space of integrals on it should be a Banach space duality between Banach spaces, with a Banach $*$-algebra being a Banach dual of the Banach space of integrals,
\item[3)] every $*$-homomorphism of $*$-algebra should be continuous in terms of Banach space norm $\n{\cdot}$ (this is assumed through the condition $\n{x^*x}=\n{x}^2$).
\end{enumerate}
Thus, as opposed to the usual algebraic approach to foundations of quantum theory, we disregard here the foundational character of $C^*$-algebras (c.f. \cite{Segal:1947:postulates}) or Jordan algebras (c.f. \cite{Emch:1972}) and associated interpretation based on spectral properties and commutative measure theory in favour of $W^*$-algebras and associated interpretation based on dualistic properties and non-commutative integration theory. On the pragmatic level, this choice can be motivated by the requirement of providing the unified treatment of commutative and non-commutative cases (see below), and by an observation that only $W^*$-algebras provide a Banach space duality between states and algebras that allows to introduce the duality between Markov maps and coarse-grainings in the infinite-dimensional non-commutative case (see next section). 

From the physical point of view, the use of non-commutative $*$-algebras instead of Hilbert spaces in the definition of information model is necessary in order to deal with the issue of unitarily inequivalent Hilbert space representations of algebras of operators, which is one of the key features of the quantum theory in infinite-dimensional (continuous) regime. This includes the continuous equilibrium statistical mechanics \cite{Emch:1972,Bratelli:Robinson:1979} and relativistic quantum field theory \cite{Haag:1992,Wald:1994,Araki:1999} as two most prominent examples. Further restriction to $W^*$-algebras is caused by the fact that $C^*$-algebras are suitable only for description of fermionic models, while the description of bosonic models necessarily requires to use the $W^*$-algebras \cite{Bratelli:Robinson:1979,Pillet:2006}.

From the statistical point of view, the use of positive parts of preduals of $W^*$-algebras instead of measure spaces $(\X,\mho(\X),\mu)$ has also an important reason. The traditional approach \cite{Fisher:1922,Kolmogorov:1933} assumes a particular choice of a background measurable space $(\X,\mho(\X))$, where $\mho(\X)$ is some particular countably additive algebra of subsets of some `sample space' $\X$ (e.g., algebra of Borel subsets of a topological space), and defines a statistical or probabilistic model as a subset of the space 
\[
	\M(\X,\mho(\X),\mu)\subseteq L_1(\X,\mho(\X),\mu)^+
\]
of all measures that are absolutely continuous with respect to a fixed measure $\mu$. However, there exist many different choices of $(\X,\mho(\X),\mu)$ such that the associated $L_1(\X,\mho(\X),\mu)$ spaces are isomorphic to the same abstract $L_1(\mho)$ space. The abstract $L_1(\mho)$ space is functorially associated to a countably additive Dedekind complete boolean (caDcb) algebra $\mho$, and the latter can be defined in terms of $(\X,\mho(\X),\mu)$ by \cite{Kakutani:1941}
\[
	\mho:=\mho(X)/\{A\in\mho(\X)\mid\mu(A)=0\}.
\]
Hence, the notion of statistical model is in fact independent of the choice of a particular sample space $\X$ and a particular measure space $(\X,\mho(\X),\mu)$. It depends only on the choice of a particular algebra $\mho$. Furthermore, the well-defined statistical models should allow to use the Steinhaus--Nikod\'{y}m isomorphism of Banach spaces 
\[
	L_1(\X,\mho(\X),\mu)^\banach\iso L_\infty(\X,\mho(\X),\mu).
\]
This requirement is crucial for considering the elements of $\M(\X,\mho(\X),\mu)$ as the Radon--Nikod\'{y}m derivatives with respect to measure $\mu$, because the latter are elements of $L_\infty(\X,\mho(\X),\mu)$. However, this isomorphism does not hold for arbitrary measure spaces \cite{Saks:1933}, but only for those $(\X,\mho(\X),\mu)$ which are localisable, what is equivalent with localisability of a pair $(\mho,\nu)$, where $\nu$ is an arbitrary semi-finite strictly positive measure on caDcb algebra $\mho$ \cite{Segal:1951}. Thus, the  definition of the statistical model which lacks any mathematical ambiguity should use the localisable pairs $(\mho,\nu)$ instead of measure spaces $(\X,\mho(\X),\mu)$. In consequence, we can restrict the considerations to the \textit{maharanisable} caDcb algebras, defined as such caDcb algebras that allow semi-finite strictly positive measures. We will call them \textit{camDcb-algebras}.

Every $W^*$-algebra is equipped with at least one normal faithful semi-finite weight $\psi$ \cite{Haagerup:1975:normal:weights}. Moreover, every commutative $W^*$-algebra $\N$ is isomorphic to $L_\infty(\mho)$ space for some camDcb-algebra $\mho$. This follows from \cite{Sakai:1971} and the fact that every localisable measure space $(\X,\mho(\X),\mu)$ generates a unique corresponding camDcb-algebra \cite{Fremlin:2000,Pietsch:2007} (the strictly positive semi-finite measure $\mu$ on $\mho$ can be derived from $\psi$, see e.g. \cite{Stratila:Zsido:1979}). But, in face of the functorial association of $L_p(\mho)$ spaces with $\mho$ \cite{Fremlin:2000}, this selects (up to isomorphism) a unique corresponding $\mho$. In such case, the space $L_1(\N)$ of integrals on $\N$ reduces to the space $L_1(\mho)$. As a result, for commutative $W^*$-algebras $\N$, the quantum information models $\M(\N)\subseteq\N_*^+$ turn into subspaces $\M(\mho)\subseteq L_1(\mho)^+$, which are precisely the well-defined statistical models.


The Falcone--Takesaki theory provides a striking generalisation of the representation independent construction of spaces of integrals to the general non-commutative case, including the generalisation of functorial association of measure-independent $L_p(\mho)$ spaces with $\mho$ to functorial association of weight-independent $L_p(\N)$ spaces with $\N$, which is based on the non-commutative flow of weights. In face of such deep structural relationships between commutative and non-commutative integration theory it is quite hard to find convincing arguments in favour of consideration of statistical theory and quantum theory as two separate theories (we do not count here the philosophical prejudices, but only the mathematical structure and the range of its applicability in concrete problems). From this perspective, relying on spectral theory, which represents quantum theoretic models in terms of commutative statistical models via the $L_2(\X,\mho(\X),\mu)$ space, cannot be longer justified. Quantum theoretic models can be quantitatively constructed and analysed as information models on its own right. Note that these conceptual changes are consequences of a single mathematical change: replacement of measure theory by integration theory.

Finally, let us note that we disregard the normalisation of states in favour of finiteness, because we consider the notion of information, quantified by finite positive integral, as more fundamental than the notion of probability, quantified by normalised measure. The quantitative evaluations $\omega\in\N_*^+$ can be used to store and transform information irrespectively of the normalisation. This perspective follows the ideas of Ingarden and Urbanik \cite{Ingarden:Urbanik:1961,Ingarden:Urbanik:1962}, but is caused also by impossibility of maintaining the relative frequency interpretation for normalised integrals on non-commutative $W^*$-algebras beyond the type II$_1$ factors, what was observed already by von Neumann \cite{vonNeumann:1937,Redei:1999}. Yet, without relative frequencies associated on the level of interpretation, the normalisation condition is just irrelevant. Thus, if one decides to consider the non-commutative integration as equally fundamental as the commutative one, there is just no mathematical or conceptual reason why one should restrict the information theoretic considerations to normalised integrals instead of finite ones.
\section{Information deviations\label{deviations.section}}
Two crucial properties that can be used to characterise the structures of information geometry (information deviations, metrics, connections) are the monotonicity under coarse graining and the generalised orthogonal decomposition under projection onto convex subspaces (that is, generalised cosine theorem). These properties encode, respectively, the requirements that ``the coarse-graining of information should always be indicated by the decrease of relative measure of information'', and ``the restriction of relative measure of information to subinformation (submodel) should be expressible in terms of its additive decomposition''.

In this section we will consider the families of deviations that are defined separately by each of these conditions. The information deviations that satisfy both of these conditions will be considered in Section \ref{entropy.section}.
\subsection{Markov monotone deviations}
In the commutative normalised case the coarse graining $T^\coa$ is defined in terms of the \textit{Markov map} $T$ by $T^\coa:\P\ni\omega\mapsto\omega\circ T\in\P$ such that, given $\omega:L_\infty(\mho,\mu;\CC)\ni f\mapsto\int\mu f\in\CC$, $T:L_\infty(\mho,\mu;\CC)\ra L_\infty(\mho',\mu';\CC)$ is a normal (monotonically continuous) positive linear map satisfying $T(1)=1$. In the non-commutative case the Markov maps are defined by normal unital completely positive linear maps $T:\N'\ra\N$ between $W^*$-algebras $\N,\N'$ \cite{Stinespring:1955}. In the finite-dimensional normalised non-commutative case it is known (see e.g. \cite{Morozova:Chentsov:1991,Petz:1994:geometry}) that there is a duality between Markov maps on $W^*$-algebras and coarse-grainings defined as normal trace-preserving completely positive maps on the corresponding spaces of density operators. However, this duality depends on a choice of underlying Hilbert space (or some reference state or weight), and the same is true for various infinite-dimensional generalisations of it. 

Using the Falcone--Takesaki theory \cite{Falcone:Takesaki:2001} and the results of Petz \cite{Petz:1984} we can show that this duality holds in the general non-commutative case without any additional assumptions. Consider a pair $(\N,\phi)$, where $\N$ is a von Neumann algebra and $\phi\in\N_*$. Define \textit{contraction} as a morphism of pairs $(\N,\phi)\ra(\N',\phi')$ given by the linear map $T:\N\ra\N'$ such that $0\leq x\leq\II\limp0\leq T(x)\leq\II$, $y\in\N^+\limp\phi'(T(y))\leq\phi(y)$, and $T$ is weakly continuous. The pairs $(\N,\phi)$ and contractions between them define a category $\Cont$. For any $T\in\Mor(\Cont)$ we define a \textit{dual map} of $T$ as such $T^\coa:\N_*'\ra\N_*$ that 
\begin{equation}
        \duality{T(x),\phi'}_{\widetilde{\N'}}=\duality{x,T^\coa(\phi')}_{\core}\;\;\forall x\in\N\;\;\forall\phi'\in\N_*',
\label{dual.cont}
\end{equation}
where $\duality{\cdot,\cdot}_{\core}$ is a Falcone--Takesaki duality (\ref{int.duality.one}). The map $T^\coa$ is a normal positive contraction, hence $(\cdot)^\coa:\Cont\ra\Cont$ is a contravariant functor such that $(\cdot)^\coa{}^\coa=\id_{\Cont}$. Moreover, following the proof in \cite{Petz:1984}, it is easy to prove that $T$ is unital and completely positive if{}f $T^\coa$ is trace-preserving and completely positive. Hence, we can replace the discussion in terms of the Markov maps (communication channels) by the discussion in terms of the dual Markov maps (coarse-grainings).

The condition of monotonicity under coarse graining imposed on the information deviation $D$ reads $D(\omega,\phi)\geq D(\omega\circ T,\phi\circ T)$, or equivalently,
\begin{equation}
        D(\omega,\phi)\geq D(T^\coa(\omega),T^\coa(\phi))
\label{monotonicity.coa}
\end{equation}
for all $\phi,\omega\in\M$ and for all coarse-grainings $T^\coa$. As remarked in introduction, in commutative finite-dimensional case (and under some mild auxiliary conditions) (\ref{monotonicity.coa}) characterises \cite{Csiszar:1978} the class of Csisz\'{a}r--Morimoto \cite{Csiszar:1963,Csiszar:1967:fdiv,Morimoto:1963} deviations. Their direct generalisation to general non-commutative case is provided by the \textit{Petz deviations} \cite{Petz:1985:quasientropies,Petz:1986:quasi:finite}
\begin{equation}
        D_\fff(\omega,\phi):=\s{\xi_\omega,\fff(\Delta_{\phi,\omega})\xi_\omega}_{\H},
\label{Petz.deviation}
\end{equation}
where $\fff:[0,\infty[\ra\RR$ is a continuous operator monotone function with $\fff(0)\geq0$ ($\fff$ is called \textit{operator monotone} if{}f $0\leq x\leq y\limp 0\leq \fff(x)\leq \fff(y)$ for any bounded operators $x,y\in\BH$). The Hilbert space $\H$ is a standard representation of the $W^*$-algebra $\N$, and $\xi_\omega$ is the unique representative of $\omega$ in the natural positive cone of the standard representation Hilbert space $\H$. All Petz deviations (\ref{Petz.deviation}) satisfy (\ref{monotonicity.coa}), but it is not known \cite{Petz:2010} whether these are the only deviations on $\N_*^+$ which have this property.
\subsection{Generalised Bregman deviations}
While the monotonicity under coarse graining requires to introduce the framework of Markov maps and the Falcone--Takesaki duality, the generalised orthogonal decomposition requires to introduce the framework of non-smooth variational analysis and the Legendre--Fenchel duality (for an exposition of the latter, see \cite{Fenchel:1949,Rockafellar:1970,Borwein:Zhu:2005,Borwein:Vanderwerff:2010}).

The existence of generalised orthogonal decompositions is involved in the representation properties of $\M$. Let $L$ and $L^\dual$ be two linear (vector) spaces over $\RR$ (or $\CC$), equipped with a bilinear duality pairing $\duality{\cdot,\cdot}_L:L\times L^\dual\ra\RR$ (or $\CC$). Let $\Psi:L\ra[-\infty,+\infty]$ and let $\dom\Psi:=\{x\in L\mid\Psi(x)<+\infty\}$. The \textit{Fenchel subdifferential} \cite{Fenchel:1949,Moreau:1963,Broensted:Rockafellar:1965} of $\Psi$ at $x\in\dom\Psi$ is defined as a set 
\begin{equation}
        \partial\Psi(x):=\{y\in L^\dual\mid\Psi(z)-\Psi(x)\geq\duality{z-x,y}_L\forall z\in L\}.
\end{equation}
The \textit{Legendre--Fenchel dual} of $\Psi$ is defined as $\Psi^\lfdual:L^\dual\ra[-\infty,+\infty]$ such that
\begin{equation}
        \Psi^\lfdual(y):=\sup_{x\in L}\{\duality{x,y}_L-\Psi(x)\}\;\;\forall y\in L^\dual.
\label{lfd}
\end{equation}
One defines $\Psi^\lfdual^\lfdual:L\ra[-\infty,+\infty]$ by $\Psi^\lfdual^\lfdual:=(\Psi^\lfdual)^\lfdual$. The functions $\Psi^\lfdual$ and $\Psi^\lfdual^\lfdual$ are convex for any $\Psi$, and $\Psi^\lfdual^\lfdual\leq\Psi$. If $\dom\Psi\neq\emptyset$, then $\Psi^\lfdual(x)>-\infty\;\forall x\in L^\dual$. If $(L,L^\dual)$ are separated locally convex vector spaces, equipped with weak-$^*$ and weak topologies, respectively, then $\Psi^\lfdual$ is weak-$^*$ lower semi-continuous, $\Psi^\lfdual^\lfdual$ is weak lower semi-continuous, and ($\Psi^\lfdual^\lfdual=\Psi$ if{}f $\Psi$ is weak lower semi-continuous and convex) \cite{Broensted:1964}. In what follows, we will always assume $\dom\Psi\neq\emptyset$. If $\Psi:L\ra\RR\cup\{+\infty\}$ is convex and $y\in L^\dual$, then the \textit{Young--Fenchel inequality} 
\begin{equation}
        \Psi(x)+\Psi^\lfdual(y)-\duality{x,y}_L\geq0
\label{Young.Fenchel.ineq}
\end{equation}
holds, with equality if{}f $y\in\partial\Psi(x)$ and $x\in\dom\Psi$. If $L$ and $L^\dual$ are over $\CC$, then $\duality{x,y}_L$ in (\ref{lfd}) and (\ref{Young.Fenchel.ineq}) is replaced by $\re\duality{x,y}_L$.

This allows to define the \textit{Bregman functional} for a convex $\Psi:L\ra\RR\cup\{+\infty\}$ as a map
\begin{equation}
        L\times L^\dual\ni(x,y)\mapsto D_\Psi(x,y):=\Psi(x)+\Psi^\lfdual(y)-\duality{x,y}_L\in[0,+\infty],
\end{equation}
if $L$ is over $\RR$, and with $\duality{x,y}_L$ replaced by $\re\duality{x,y}_L$ if $L$ is over $\CC$. By definition, $D_\Psi$ is convex in each variable separately, $D_\Psi(x,y)\geq0\;\forall(x,y)\in L\times L^\dual$, and $D_\Psi(x,y)=0\iff y\in\partial\Psi(x)$. We will call Bregman functional $D_\Psi$ \textit{adequate for $V\subseteq L$} if{}f $\partial\Psi(x)\neq\emptyset\;\forall x\in\dom\Psi\cap V$. There exist various criteria for non-emptiness of Fenchel subdifferential which can be used to ensure adequacy of $D_\Psi$. In particular, if $(L,L^\dual)$ are Banach spaces, and $\Psi$ is convex and lower semi-continuous, then the Fenchel--Rockafellar theorem \cite{Rockafellar:1970,Borwein:Zhu:2005} states that $\partial\Psi(x)\neq\emptyset\;\forall x\in\CORE(\dom\Psi)$, where $x\in\CORE(X)$ for $\emptyset\neq X\subseteq L$ if{}f $\forall h\in S_L\;\;\exists \varepsilon >0\;\;\forall t\in[0,\varepsilon]\;\;x+th\in X$, with $S_L$ denoting a unit sphere in $L$. 

Define \textit{dual coordinate system} (or \textit{dual representation}) as a map $(r,s):\M\times\M\ni(\omega,\phi)\mapsto(r(\omega),s(\phi))\in L\times L^\dual$. We define the \textit{generalised Bregman deviation} as a function
\begin{equation}
        D_\Psi:\M\times\M\ni(\omega,\phi)\mapsto D_\Psi(\omega,\phi):=D_\Psi(r(\omega),s(\phi))\in[0,+\infty],
\label{gbd}
\end{equation}
where $(r,s)$ is a dual coordinate system such that $r(\omega)\in\partial\Psi(r(\omega))$ $\forall\omega\in\M$, while $D_\Psi$ is a Bregman functional adequate for $\cod(r)$. By definition, $D_\Psi(\omega,\phi)$ is convex in each variable separately, $D_\Psi(\omega,\phi)\geq0\;\forall\omega,\phi\in\M$, and $\omega=\phi\limp D_\Psi(\omega,\phi)=0$ $\forall\omega\in\M$. This weakening of the usual property of deviation ($\omega=\phi\iff D(\omega,\phi)=0$) to one-sided implication is the price paid for defining $D_\Psi$ in terms of a weakly constrained (hence, quite general) variational problem. In order to impose an implication in opposite direction, one would have to require additional conditions that are not natural at this level of generality (they will be discussed below). Note also that due to adequacy of $D_\Psi$, the condition $s(\omega)\in\partial\Psi(r(\omega))$ is well defined for all $\omega\in\M$. It can be understood either as a condition on allowed dual coordinate systems if $\Psi$ is given, or as a condition on $\Psi$ if $(r,s)$ is given. To summarise, an information deviation $D:\M\times\M\ra[0,\infty]$ is a generalised Bregman deviation, denoted $D_\Psi$, if{}f there exists a pair of dualised vector spaces $(L,L^\dual,\duality{\cdot,\cdot}_L)$, a pair of functions $r:\M\ra L$ and $s:\M\ra L^\dual$ and a convex function $\Psi:L\ra\RR\cup\{+\infty\}$ satisfying the conditions
\begin{equation}
 \partial\Psi(x)\neq\emptyset\;\forall x\in\dom\Psi\cap\cod r,\;\;\;
                s(\omega)\in\partial\Psi(r(\omega))\;\forall\omega\in\M,
\end{equation}
and such that
\begin{equation}
                D(\omega,\phi)=\sup_{x\in L}\{\duality{x,s(\phi)}_L-\Psi(x)\}-\duality{r(\omega),s(\phi)}_L-\Psi(r(\omega)),
\label{bdk}
\end{equation}
with $\duality{\cdot,\cdot}_L$ replaced by $\re\duality{\cdot,\cdot}_L$ if $L$ is over $\CC$. Note that one could also define the functional
\begin{equation}
                \hat{D}_\Psi(\omega,\phi)=\sup_{\omega\in\M}\{\duality{r(\omega),s(\phi)}_L-\Psi(r(\omega))\}-\duality{r(\omega),s(\phi)}_L-\Psi(r(\omega)),
\label{bdk.bad}
\end{equation}
but as long as $\cod(r)\subsetneq\dom\Psi\cap L$, (\ref{bdk.bad}) is essentially different from (\ref{bdk}), due to non-linear character of the $\sup$ operation. 
\subsection{Dualisers of representation\label{dualisers.section}}
The generalised Bregman deviation (\ref{bdk}) exposes the underlying dualistic and variational properties of the deviation functional. However, often the other representation of Bregman deviation is used, which exposes its geometric properties at the price of the non-trivial restrictions on the domain of duality and convexity. Usually these restrictions are imposed in order to adapt to \textit{presupposed} topological framework and some presupposed form of representation of Bregman deviation, and the goal is to show that such Bregman deviation encodes the Legendre case of the Legendre--Fenchel duality with the dual variable $y\in L^\dual$ given by some suitably defined notion of derivative (e.g. Fr\'{e}chet, G\^{a}teaux, one-sided G\^{a}teaux), see e.g. \cite{Censor:Lent:1981,Bauschke:Borwein:1997,Butnariu:Resmerita:2006,Borwein:Vanderwerff:2010} for standard examples in commutative case and \cite{Petz:2007:Bregman} for an example in finite-dimensional non-commutative case. Our approach is different, because we do not assume any fixed framework for continuity or smoothness, so we can analyse the general relationship between explicitly dual (generalised) Bregman deviation and its \textit{standard} (hence, restricted) version, which has both arguments represented on the same space. We will construct now a general framework for conversion between these two forms of the Bregman deviation, which is independent of any particular assumptions about continuity or differentiability. The key role in this setting will be played by the (generally, non-linear) \textit{dualiser} function. It is an infinite-dimensional generalisation of the transformation between the dual coordinate systems of Bregman functional, which in the finite-dimensional case is usually encoded in terms of derivative.

For a convex $\Psi:L\ra\RR\cup\{+\infty\}$ define a function $f_\Psi:L\ra L^\dual$ such that there exists a set $\emptyset\neq V\subseteq L$ satisfying:
\begin{enumerate}
\item[($1_f$)] $f_\Psi:V\ra f_\Psi(V)$ is an isomorphism,
\item[($2_f$)] $\forall y\in V\;\;\Psi^\lfdual(f_\Psi(y))-\Psi(y)=\duality{y,f_\Psi(y)}_L$,
\item[($3_f$)] $\forall x\in\dom\Psi\cap V\;\;\exists!z\in f_\Psi(V)\;\;z\in\partial\Psi(x)$.
\end{enumerate}
Such function $f_\Psi$ will be called a \textit{dualiser}, while the set $V$ will be denoted $\admf$ and called and \textit{admissible domain} of $f_\Psi$. We will consider $f_\Psi$ only on the set $\admf$, so we define $\cod f_\Psi:=f_\Psi(\admf)$. The function $\Psi$ will be called \textit{dualisable} with respect to $(L,L^\dual,\duality{\cdot,\cdot}_L)$ if{}f there exists at least one dualiser $f_\Psi$. Note that dualisability of $\Psi$ is a relative property: a change of domain $L$ or a change of a duality structure $\duality{\cdot,\cdot}_L$ on $L$ changes the available dualisers.
Note also that the condition ($3_f$) can be equivalently written as
\begin{equation}
        \cod f_\Psi\cap\partial\Psi|_{\admf\cap\dom\Psi}(x)=\{*\},
\label{condition.three.f}
\end{equation}
where $\{*\}$ is a singleton. For convenience of notation, till the end of this section we will assume that $L$ is over $\CC$ (if it is over $\RR$, then one has just to drop `$\re$' everywhere).

If $D_\Psi$ is a Bregman functional and $\Psi$ is dualisable with a dualiser $f_\Psi$, then the \textit{unbounded standard Bregman functional} is defined as a map
\begin{equation}
\bar{D}_\Psi:L\times\admf\ni(x,y)\mapsto\bar{D}_\Psi(x,y):=\Psi(x)-\Psi(y)-\re\duality{x-y,f_\Psi(x)}_L\in\RR\cup\{+\infty\},
\end{equation}
while the \textit{bounded standard Bregman functional} is defined as a map
\begin{equation}
\bar{D}_\Psi:\dom\Psi\times\admf\ni(x,y)\mapsto\bar{D}_\Psi(x,y):=\Psi(x)-\Psi(y)-\re\duality{x-y,f_\Psi(x)}_L\in\RR.
\end{equation}
By definition, $\bar{D}_\Psi$ satisfies
\begin{enumerate}
\item[i)] $\bar{D}_\Psi(x,y)=D_\Psi(x,f_\Psi(y))$,
\item[ii)] $\bar{D}_\Psi(x,y)\geq0$,
\item[iii)] $\bar{D}_\Psi(x,y)=0\iff x=y$,
\end{enumerate}
for all $(x,y)\in L\times\admf$ (or for all $(x,y)\in\dom\Psi\times\admf$, if $\bar{D}_\Psi$ is bounded). The two-sided implication appears here at the price of loss of convexity of $\bar{D}_\Psi$ in the second variable (what is a common problem in standard treatments, see e.g. \cite{Bauschke:Borwein:2001}). This is because (the inverse of) a dualiser $f_\Psi$ may not preserve the convexity properties.

The \textit{(standard) Bregman deviation} is defined correspondingly as
\begin{equation}
        \bar{D}_\Psi(\omega,\phi):=\bar{D}_\Psi(r(\omega),f^{-1}_\Psi\circ s(\phi))=\Psi(r(\omega))-\Psi(f^{-1}_\Psi\circ s(\phi))-\re\duality{r(\omega)-f^{-1}_\Psi\circ s(\phi),s(\phi)}_L,
\label{std.Breg.dev}
\end{equation}
where $(r,s):\M\times\M\ra L\times L^\dual$ is a dual coordinate system such that $s(\omega)\in\partial\Psi(r(\omega))\;\forall\omega\in\M$. By definition, $\bar{D}_\Psi(\omega,\phi)=D_\Psi(\omega,\phi)$, where $D_\Psi$ is a Bregman functional adequate for $\cod(r)$. It follows that a single generalised Bregman deviation can have several different representations in terms of standard Bregman deviations, depending on the choice of dualiser $f_\Psi:L\supseteq\admf\ra\cod f_\Psi\subseteq L^\dual$. If $\bar{D}_{\Psi,f_1}$ and $\bar{D}_{\Psi,f_2}$ are two standard Bregman functionals defined from a single generalised Bregman functional $D_\Psi$ by two dualisers $f_1$ and $f_2$, then they are equal to each other on $V\subseteq\adm f_1\cap\adm f_2$ if{}f there exists a dualiser $f_3$ of $\Psi$ such that $\adm f_3=V$. The existence of different dualisers is equivalent to $\partial\Psi(r(\cdot))$ being a non-singleton, non-empty, set-valued function. From the geometric perspective, the choice of a particular dualiser (and an associated representation of generalised Bregman deviation in terms of standard Bregman deviation) corresponds to choice of a particular section of the presheaf $\M\ni\omega\mapsto\partial\Psi(r(\omega))\subseteq L^\dual$. 

The definition of bounded standard Bregman functional $\bar{D}_\Psi$ implies that $\bar{D}_\Psi$ satisfies the \textit{generalised cosine equation}
\begin{equation}
        \bar{D}_\Psi(r_1,r_2)+\bar{D}_\Psi(r_2,r_3)-\bar{D}_\Psi(r_1,r_3)=\re\duality{r_1-r_2,f_\Psi(r_3)-f_\Psi(r_2)}_L\;\;\forall r_1,r_2,r_3\in\admf\cap\dom\Psi.
\label{generalised.cosine}
\end{equation}
Under suitable assumptions that guarantee existence and uniqueness of the solution of the corresponding variational problem, this equation can be turned into a theorem on existence and uniqueness of generalised additive decomposition of information deviation under projection onto subspace (submodel).

Let $y\in\admf\cap\dom\Psi$, let $R\subseteq\admf\cap\dom\Psi$ be non-empty and convex, and let $r\in R$. The element $\bar{x}\in R$ will be called the \textit{Bregman projection} of $y$ on $R$, and denoted by $P_R^\Psi(y)$, if{}f
\begin{equation}
        \bar{x}=\arg\inf_{r\in R}\{\bar{D}_\Psi(r,y)\}.
\label{Bregman.projection}
\end{equation}
The problem with this definition is that in general case $P^\Psi_R(y)$ might not exist or might be non-unique. The existence and uniqueness of Bregman projection can be guaranteed under various assumptions. In particular, the existence can be guaranteed by means of Bauer's theorem \cite{Bauer:1958} (if $L$ is a locally convex space, $R$ is weakly compact, and $\bar{D}_\Psi$ is weakly lower semi-continuous). On the other hand, if $L$ is a reflexive Banach space, $R$ is closed, $\bar{D}_\Psi$ is lower semi-continuous, strictly convex, and G\^{a}teaux differentiable at $y$, with $\INT\dom\bar{D}_\Psi\neq\emptyset$, $R\cap\dom\bar{D}_\Psi$ and $y\in\INT\dom\bar{D}_\Psi$, then $P_R^\Psi(y)$ is at most a singleton \cite{Borwein:Vanderwerff:2010}. Unfortunately, we were unable to find the sufficient conditions for uniqueness and existence of the Bregman projections that would be phrased without appealing to particular topological framework. Expressing such conditions in general terms of dualisers would be a valuable result. 

Let $(L,L^\dual)$ be a dual pair of separated locally convex spaces, let $R\subseteq\admf\cap\dom\Psi$ be a non-empty convex, affine, weakly closed set, and let $\bar{x}:=P_R^\Psi(y)$ be a unique Bregman projection of $y\in\admf\cap\dom\Psi$. Then
the \textit{generalised pythagorean theorem} 
\begin{equation}
        \bar{D}_\Psi(r,\bar{x})+\bar{D}_\Psi(\bar{x},y)=\bar{D}_\Psi(r,y)\;\;\;\;\forall r\in R
\label{generalised.pythagorean}
\end{equation}
holds if{}f the orthogonality condition $\re\duality{r-\bar{x},f_\Psi(y)-f_\Psi(\bar{x})}_L=0$ is satisfied. Such $\bar{x}=P^\Psi_R(y)$ is called a \textit{(dually) orthogonal Bregman projection}. The property (\ref{generalised.pythagorean}) generalises the additive decompositions of norm under linear projections on closed convex subsets in the Hilbert space to the class of non-linear Bregman projections on closed convex subsets in the linear space $L$. Note that the `orthogonality' of projection is understood in the sense of the linear duality pairing $\duality{\cdot,\cdot}_L$, while the non-linearity of projection $P^\Psi_R$ corresponds to the non-linear dualiser $f_\Psi$.
\section{The $\gamma$-family of relative entropies\label{entropy.section}}
\subsection{Families of $\gamma$-deviations}
By imposing monotonicity under coarse-graining condition on the generalised Bregman deviation (or on a corresponding standard Bregman deviation) one obtains strong restriction on the allowed form of the deviation and corresponding dual coordinate systems (representations). For the finite-dimensional commutative case, Amari \cite{Amari:2009:alpha:divergence} has shown that the unique deviation that satisfies these constraints is the Zhu--Rohwer $\gamma$-deviation \cite{Zhu:Rohwer:1995,Zhu:Rohwer:1997,Zhu:Rohwer:1998} 
\begin{equation}
        D_\gamma(\omega,\phi)=
        \int\frac{
                \gamma\mu_\omega+
                (1-\gamma)\nu_\phi-
                \mu_\omega^\gamma\nu_\phi^{1-\gamma}
                }
                {\gamma(1-\gamma)},
\label{ZhuRohwer.deviation}
\end{equation}
where $\mu_\omega$ and $\nu_\phi$ are the finite positive measures corresponding to the Daniell--Stone integrals $\omega$ and $\phi$, while $\mu_\omega^\gamma\nu^{-\gamma}_\phi=(\frac{\mu_\omega}{\nu_\phi})^\gamma$ is just the $\gamma$-th power of the Radon--Nikod\'{y}m derivative $\frac{\mu_\omega}{\nu_\phi}$. The most general deviation satisfying both constraints in the non-commutative case that has been known so far is the Jen\v{c}ov\'{a}--Ojima $\gamma$-deviation \cite{Jencova:2003:arXiv,Jencova:2005,Ojima:2004}
\begin{equation}
        D_\gamma(\omega,\phi)=
        \frac{
                \gamma\omega(\II)+
                (1-\gamma)\phi(\II)-
                \re\duality{
                        u\Delta^\gamma_{\omega,\psi},
                        v\Delta_{\phi,\psi}^{1-\gamma}
                        }_\psi
                }
                {\gamma(1-\gamma)},
\label{JencovaOjima.deviation}
\end{equation}
where $\psi\in\W_0(\N)$ is an arbitrary reference functional, $\duality{\cdot,\cdot}_\psi$ is the Banach space duality pairing between the Araki--Masuda non-commutative $L_{1/\gamma}(\N,\psi)$ and $L_{1/(1-\gamma)}(\N,\psi)$ spaces, $\Delta_{\phi,\psi}$ is a relative modular operator, while $u$ (resp., $v$) is an operator arising from the polar decomposition of $\omega$ (resp., $\phi$), $\omega(xu)=\duality{u\Delta_{\omega,\psi},x^\dual}_\psi$ $\forall x\in\N$, where $x^\dual$ denotes a dual of $x$ with respect to the Araki--Masuda duality $\duality{\cdot,\cdot}_\psi$. 

The deviations given by (\ref{ZhuRohwer.deviation}) and (\ref{JencovaOjima.deviation}) are well defined for $\gamma\in\RR\setminus\{0,1\}$, however the characterisation results include also the boundary case $\gamma\in\{0,1\}$, for which the corresponding deviations are derived from $D_\gamma$ by passing to the limit with $\gamma$ under integral sign. If the normalisation condition is imposed, then the uniqueness results are stronger, selecting precisely the boundary functionals: the Kullback--Leibler \cite{Kullback:Leibler:1951,Kullback:1959} deviation
\begin{equation}        D_0(\omega,\phi)=D_1(\phi,\omega)=\int\nu_\phi\log\frac{\nu_\phi}{\mu_\omega}
\label{Kullback.Leibler}
\end{equation}
in the finite-dimensional commutative case \cite{Csiszar:1991}, and the Umegaki--Araki \cite{Umegaki:1962,Araki:1976:relative:entropy:I,Araki:1977:relative:entropy:II} deviation 
\begin{equation}
        D_0(\omega,\phi)=D_1(\phi,\omega)=\tr(\rho_\phi(\log\rho_\phi-\log\rho_\omega))=\tr\left(\rho_\phi^{1/2}(\log\Delta_{\phi,\omega})\rho_\phi^{1/2}\right)
\label{Umegaki.Araki}
\end{equation}
in the finite-dimensional non-commutative case \cite{Petz:2007:Bregman}.

Despite these similarities, the Jen\v{c}ov\'{a}--Ojima deviation is not a canonical non-commu\-ta\-tive generalisation of the Zhu--Rohwer deviation. The construction of the former is dependent on the choice of fixed reference weight $\psi$, while the latter does not depend on any additional measure. (Nevertheless, the values taken by the Jen\v{c}ov\'{a}--Ojima deviation are independent of the choice of reference $\psi$.) In what follows, we will use the Falcone--Takesaki theory in order to introduce a new family of information deviations that is a proper non-commutative generalisation of the Zhu--Rohwer deviations and is also a proper generalisation of the Jen\v{c}ov\'{a}--Ojima deviations.
\subsection{General form of family of $\gamma$-deviations}
Consider the \textit{$\gamma$-embedding} (\textit{$\gamma$-coordinate}) functions on $\N_*^+$ valued in $L_{1/\gamma}(\N)$ spaces:
\begin{equation}
        \ell_\gamma:\N_*^+\ni\omega\mapsto\ell_\gamma(\omega):=\frac{\omega^\gamma}{\gamma}\in L_{1/\gamma}(\N),
\label{ell.gamma}
\end{equation}
with $\gamma\in]0,1]$. By definition, the embeddings $\ell_\gamma$ and $\ell_{1-\gamma}$ encode the duality (\ref{dual.pairing.Lp.FT}),
\begin{equation}
        \N_*^+\times\N_*^+\ni(\omega,\phi)\mapsto\int\ell_\gamma(\omega)\ell_{1-\gamma}(\phi)\in\CC.
\end{equation}
The special case of these maps were introduced by Nagaoka and Amari \cite{Nagaoka:Amari:1982} as coordinate systems in normalised commutative finite-dimensional setting, and since then they became a standard tool of information geometry theory. However, the Nagaoka--Amari formulation, as well as all its later applications, including the non-commutative generalisations, is based on using the $\gamma$-powers of densities (Radon--Nikod\'{y}m derivatives or Pedersen--Takesaki operator densities) with respect to a fixed reference measure, state, or weight. An important attempt to circumvent this problem in the commutative case was made by Zhu \cite{Zhu:1998:lebesgue,Zhu:Rohwer:1998}, who considered the spaces of measures constructed through an equivalence relation based on $\gamma$-powers of Radon--Nikod\'{y}m derivatives, but without fixing any particular reference measure. However, his work remained unfinished and widely unknown, and it covered only the commutative case. The embeddings (\ref{ell.gamma}) solve these problems.

Due to the Falcone--Takesaki theory we are able to make the reference-independent approach strict and valid in all cases, including the infinite-dimensional non-commutative one. Let us define \cite{Kostecki:2011:AIP} the general form of \textit{quantum $\gamma$-deviation} on $\N_*^+$, $D_\gamma:\N_*^+\times\N_*^+\ni(\omega,\phi)\mapsto D_\gamma(\omega,\phi)\in\RR$, by
\begin{equation}
                D_\gamma(\omega,\phi):=\int\left(\frac{\omega}{1-\gamma}+\frac{\phi}{\gamma}-\re\left(\ell_\gamma(\omega)\ell_{1-\gamma}(\phi)\right)\right)=\re\int\left(\frac{\gamma\omega+(1-\gamma)\phi-\omega^\gamma\phi^{1-\gamma}}{\gamma(1-\gamma)}\right),
\label{rpk.deviation}   
\end{equation}
where $\gamma\in]0,1[$. This definition can be extended to include the boundary values $\gamma\in\{0,1\}$, by
\begin{equation}
        D_1(\phi,\omega):=D_0(\omega,\phi):=\int\lim_{\gamma\ra0}\left(\frac{\omega}{1-\gamma}+\frac{\phi}{\gamma}-\re\left(\ell_\gamma(\omega)\ell_{1-\gamma}(\phi)\right)\right).
        \label{rpk.deviation.boundary}
\end{equation}
The deviation $D_\gamma$ reduces to the following special cases: 
\begin{itemize}
\item the Jen\v{c}ov\'{a}--Ojima $\gamma$-deviation (\ref{JencovaOjima.deviation}) when the choice of a reference state $\psi\in\N_{*0}^+$, or a reference weight $\psi\in\W_0(\N)$, for the isometric isomorphism with the Araki--Masuda $L_p(\N,\psi)$ space is provided,
\item the Umegaki--Araki deviation (\ref{Umegaki.Araki}) in the representation-independent Petz's form \cite{Petz:1985:properties} 
\begin{equation}
D_0(\omega,\phi)=D_1(\phi,\omega)=i\lim_{t\ra 0}\frac{\phi}{t}\left(\left(\frac{\DD\omega}{\DD\phi}\right)_t-\II\right),
\end{equation}
for $\n{\phi}=\n{\omega}=1$, $\supp(\omega)\geq\supp(\phi)$, and $\gamma\ra 0$ limit, where $\supp(\omega)$ denotes the support projection of $\omega$,
\item the Hasegawa $\gamma$-deviation \cite{Hasegawa:1993} 
\begin{equation}
D_\gamma(\omega,\phi)=\frac{\psi(\rho_\omega-\rho_\omega^\gamma\rho_\phi^{1-\gamma})}{\gamma(1-\gamma)},
\end{equation}
for semi-finite $\N$, and $\rho_\phi$ and $\rho_\omega$ defined as, respectively, the Pedersen--Takesaki densities of normalised states $\phi$ and $\omega$ with respect to a faithful normal semi-finite trace $\psi$ that acts on a standard representation of $\N\iso L_\infty(\N)$ on $\H\iso L_2(\N)$. In this case the Pedersen--Takesaki densities reduce to the Dye--Segal densities \cite{Dye:1952,Segal:1953}, hence $\phi(\cdot)=\psi(\rho_\phi\;\cdot)$ and $\omega(\cdot)=\psi(\rho_\omega\;\cdot)$. If the standard representation of $\N$ on $\H$ is isomorphic to $\BH$ as a von Neumann algebra and trace $\psi$ is normalised ($\psi(\II)=1$), then $\psi(\cdot)=\tr(\cdot)$, where $\tr$ is a standard trace on $\BH$,
\item the Zhu--Rohwer $\gamma$-deviation (\ref{ZhuRohwer.deviation}) for commutative $\N$, and $\nu_\phi$ absolutely continuous with respect to $\mu_\omega$,
\item the Kullback--Leibler deviation (\ref{Kullback.Leibler}) for commutative $\N$, $\nu_\phi$ absolutely continuous with respect to $\mu_\omega$, $\int\mu_\omega=1=\int\nu_\phi$, and $\gamma\ra0$ limit,
\item the Cressie--Read--Amari $\gamma$-deviation\footnote{The $\gamma$-deviation family (\ref{cressie.read.amari}) corresponds bijectively, but is not equal, to the $\gamma$-deviation families of Chernoff \cite{Chernoff:1952}, R\'{e}nyi \cite{Renyi:1961}, P\'{e}rez \cite{Perez:1967}, Havrda--Ch\'{a}rvat \cite{Havrda:Charvat:1967}, and Tsallis \cite{Tsallis:1988} (for a review with calculations, see e.g. \cite{Crooks:Sivak:2011}).} \cite{Cressie:Read:1984,Amari:1985,Liese:Vajda:1987,Read:Cressie:1988}
\begin{equation}
	D_\gamma(\omega,\phi)=\frac{1}{\gamma(1-\gamma)}\int\upsilon(p_\omega-p_\omega^\gamma q_\phi^{1-\gamma}),
\label{cressie.read.amari}
\end{equation}
for semi-finite commutative $\N$, and normalised measures $\nu_\phi$ and $\mu_\omega$ ($\int\mu_\omega=1=\int\nu_\phi$) that are absolutely continuous with respect to a strictly positive semi-finite measure $\upsilon$ on the countably additive Dedekind complete boolean algebra $\mho$ derived from $\N$. The functions $q_\phi$ and $p_\omega$ are defined, respectively, as the Radon--Nikod\'{y}m derivatives of $\nu_\phi$ and $\mu_\omega$ with respect to $\upsilon$: $\phi(\cdot)=\int\nu_\phi(\cdot)=\int\upsilon (q_\phi\;\cdot)$, and $\omega(\cdot)=\int\mu_\omega(\cdot)=\int\upsilon (p_\omega\;\cdot)$.
\end{itemize}
It is quite remarkable that the mathematical form of $\gamma$-deviation given by (\ref{rpk.deviation}) resembles so strong similarity with its commutative special case (\ref{ZhuRohwer.deviation}). 
\subsection{Properties of $\gamma$-deviations}
Let us turn to the discussion of the properties of the family (\ref{rpk.deviation}). A direct calculation shows that (\ref{rpk.deviation}) is a generalised Bregman deviation (\ref{bdk}) with 
\begin{equation}
        \Psi_\gamma(x)=\frac{1}{1-\gamma}\int(\gamma x)^{1/\gamma}=\frac{1}{1-\gamma}\n{\gamma x}_{1/\gamma}^{1/\gamma},
\end{equation}
and $\Psi_\gamma(\ell_\gamma(\omega))=\frac{1}{1-\gamma}\omega(\II)$. This deviation has a dualiser $f_{\Psi_\gamma}(x)=\ell_{1-\gamma}\circ\ell_\gamma^{-1}(x)$, which is a homeomorphism $f_{\Psi_\gamma}:L_{1/\gamma}(\N)\ra L_{1/(1-\gamma)}(\N)$. Thus, $D_\gamma$ can be written in terms of the generalised Bregman deviation (\ref{bdk}), which takes the form
\begin{equation}
        D_{\Psi_\gamma}(\omega,\phi)=\Psi_\gamma(\ell_\gamma(\omega))+\Psi_{1-\gamma}(\ell_{1-\gamma}(\phi))-\re\duality{\ell_\gamma(\omega),\ell_{1-\gamma}(\phi)}_\core.
\label{rpk.gamma.std.Breg}
\end{equation}

The straightforward calculation based on the definitions (\ref{rpk.deviation}) and (\ref{ell.gamma}) shows that $D_\gamma$ satisfies the generalised cosine equation
\begin{equation}
	D_\gamma(\omega,\phi)+D_\gamma(\phi,\psi)=D_\gamma(\omega,\psi)+\re\int\left(\ell_\gamma(\omega)-\ell_\gamma(\phi)\right)\left(\ell_{1-\gamma}(\psi)-\ell_{1-\gamma}(\phi)\right),
\label{gen.cosine.rpk.dev}
\end{equation}
as well as 
\begin{equation}
        D_\gamma(\omega,\phi)=D_{1-\gamma}(\phi,\omega).
\label{rpk.dev.index.duality}
\end{equation}
The equation (\ref{gen.cosine.rpk.dev}) turns into the `ordinary' generalised cosine equation (\ref{generalised.cosine}) for the corresponding standard Bregman deviation (\ref{rpk.gamma.std.Breg}), while the equation (\ref{rpk.dev.index.duality}) turns into the `representation-index duality' equation
\begin{equation}
        \bar{D}_{\Psi_\gamma}(y,x)=\bar{D}_{\Psi_{1-\gamma}}(f_{\Psi_\gamma}(x),f_{\Psi_\gamma}(y)),
\label{rep.idx.duality.eq}
\end{equation}
where $x,y\in L_{1/\gamma}(\N)$. The finite-dimensional commutative version of the equation (\ref{rep.idx.duality.eq}), with a dualiser given by derivative, was discussed in \cite{Zhang:2004:divergence}.

Furthermore, from the uniform convexity and the uniform smoothness of $L_{1/\gamma}(\N)$ spaces it follows that the norm $\n{\cdot}_{1/\gamma}$ is Fr\'{e}chet differentiable at all points $x\in L_{1/\gamma}(\N)$ except $x=0$. The Fr\'{e}chet derivative of the norm allows to calculate the corresponding Fr\'{e}chet derivative of the function $\Psi_\gamma(x)$ in the direction $y$, which reads
\begin{equation}
        \DDD_F(\Psi_\gamma(x))(y)=\re\duality{y,f_{\Psi_\gamma}(x)}_\core=\re\duality{y,\ell_{1-\gamma}\circ\ell^{-1}_\gamma(x)}_\core.
\end{equation}
For $\gamma<1$ the function $\Psi_\gamma(x)$ is Fr\'{e}chet differentiable also at $x=0$, with 
\begin{equation}
	\DDD_F(\Psi_\gamma(0))(y)=\re\duality{y,f_\Psi(0)}_\core=0.
\end{equation}
This gives a differential formula for the standard Bregman functional $\bar{D}_{\Psi_\gamma}$,
\begin{equation}
        \bar{D}_{\Psi_\gamma}(y,x)=\Psi_\gamma(y)+\Psi_{1-\gamma}(\ell_{1-\gamma}\circ\ell_\gamma^{-1}(x))-\DDD_F(\Psi_\gamma(x))(y).
\end{equation}
The uniform smoothness and uniform convexity of $L_p(\N)$ allows to strictly follow Jen\v{c}ov\'{a}'s proof of Proposition~6.1.(i)-(ii) in \cite{Jencova:2005}, what results in a proof that, for every $\phi\in\N_*^+$,
\begin{enumerate}
\item[1)] $D_\gamma(\omega,\phi)\geq0$,
\item[2)] $D_\gamma(\omega,\phi)=0\iff\omega=\phi$.
\end{enumerate}

As stated above, for any choice of a reference weight $\psi\in\W_0(\N)$, $D_\gamma$ turns into the Jen\v{c}ov\'{a}--Ojima deviation. The latter is in turn the particular case of Petz's deviation (\ref{Petz.deviation}) with the operator convex function $\fff$ given by
\begin{equation}
        \fff_\gamma(t)=\frac{1}{\gamma}+\frac{1}{1-\gamma}t-\frac{1}{\gamma(1-\gamma)}t^\gamma,
\label{g.gamma}
\end{equation}
see \cite{Jencova:2005} for a proof. The identification of $D_\gamma$ as a member of Petz's family with the function (\ref{g.gamma}) can be also obtained more directly, without any reference to the Araki--Masuda $L_p(\N,\psi)$ spaces. This is provided using the identification (\ref{FT.Hilbert.space}) of $L_2(\N)$ with the standard representation Hilbert space $\H$, which gives
\begin{equation}
        \int\omega^\gamma\phi^{1-\gamma}=\s{\xi_\phi,\left(\frac{\DD\omega}{\DD\phi}\right)_{-i\gamma}\xi_\phi}_\H=\s{\xi_\phi,\Delta^\gamma_{\omega,\phi}\xi_\phi}_\H,
\end{equation}
where $\xi_\phi$ is a representative of $\phi$ in the standard cone $L_2(\N)^+$ of the standard representation Hilbert space $\H$. The last equality follows from the fact that $\left(\frac{\DD\omega}{\DD\phi}\right)_z\xi_\phi$ agrees with $\Delta^{iz}_{\omega,\phi}\xi_\phi$ for $z\in\RR$ \cite{Takesaki:2003}. From the fact that $D_\gamma$ is a Petz deviation it follows that it satisfies the following properties \cite{Petz:1985:quasientropies,Jencova:2005}:
\begin{enumerate}
\item[1)] $D_\gamma(\omega,\phi)\geq D_\gamma(T^\coa\circ\omega,T^\coa\circ\omega)$,
\item[2)] $D_\gamma$ is jointly convex on $\N_*^+\times\N_*^+$,
\item[3)] $D_\gamma$ is lower semi-continuous on $\N_*^+\times\N_{*0}^+$ endowed with the product of norm topologies.
\end{enumerate}

Using the lower semi-continuity and convexity properties of $D_\gamma$ together with the fact that $L_{1/\gamma}(\N)$ spaces are Banach spaces, one can take the advantage of the theorems on existence and uniqueness of the solutions of the variational problems for the convex lower semi-continuous functionals in the Banach space setting. In particular, if $\psi\in\N_*^+$, $C\subseteq\N_*^+$, $\phi\in C$, and $\ell_\gamma(C)$ is a convex set, then the following conditions are equivalent:
\begin{enumerate}
\item[i)] $D_\gamma(\phi,\psi)=\inf_{\varphi\in C}\{D_\gamma(\varphi,\psi)\}$,
\item[ii)] $D_\gamma(\omega,\psi)\geq D_\gamma(\phi,\psi)+D_{1-\gamma}(\phi,\omega)$.
\end{enumerate}
Moreover, if such $P_C^{\Psi_\gamma}(\psi)=\phi$ exists, then it is unique. This condition is satisfied if $C$ is closed in the topology induced by the $\gamma$-embedding of the weak topology of $\N_*^+$ in $L_p(\N)$. The explicit proof of these statements can be obtained by repeating the proofs of Propositions 7.1, 8.1 and 8.2 of \cite{Jencova:2005} with the obvious replacement of $L_{1/\gamma}(\N,\psi)$ by $L_{1/\gamma}(\N)$.

The $\gamma$-family of deviation functionals defined by (\ref{rpk.deviation}) provides a proper infinite-dimensional non-commutative generalisation of the $\gamma$-families of  Zhu--Rohwer deviations and Jen\v{c}ov\'{a}--Ojima deviations that is defined directly in terms of canonical non-commu\-ta\-ti\-ve $L_{1/\gamma}(\N)$ spaces and belongs to an intersection of the generalised Bregman deviations and Petz's deviations. The canonical character of the structures employed in the construction of $D_\gamma$, and the recent result of Amari \cite{Amari:2009:alpha:divergence} for Zhu--Rohwer deviation in finite-dimensional commutative case, leads us to state the following:

\ \\
\textbf{Conjecture $\clubsuit$}. The family $D_\gamma(\omega,\phi)$, $\gamma\in[0,1]$, defined by (\ref{rpk.deviation}) and (\ref{rpk.deviation.boundary}) is the unique family of deviation functionals on $\N^+_*$ satisfying (\ref{monotonicity.coa}) and (\ref{bdk}).

\ \\
Further properties of $D_\gamma$ family, the issue of selection of a particular $\gamma\in[0,1]$, as well as the extension of other structures of quantum information geometry in infinite-dimensional non-commutative case based on Falcone--Takesaki theory will be the main topic of the next paper \cite{Kostecki:2011:QIG}.

As a final remark, let us note that for $\gamma=1/2$ the $L_{1/\gamma}(\N)$ space turns into a Hilbert space $\H$, while the standard Bregman functional $\bar{D}_\gamma$ associated with $D_\gamma$ turns into the norm distance
\begin{equation}
        \bar{D}_{\Psi_{1/2}}(x,y)=\frac{1}{2}\n{x-y}^2_\H.
\end{equation}
Hence, the Bregman projection of information geometric theory  based on $\gamma$-deviations (\ref{rpk.deviation}) turns in this particular case to the Hilbert space norm minimisation. This shows that the framework of Hilbert space geometry is just a special case of much more general framework of information geometry of spaces of non-commutative integrals. From this perspective, the traditional application of spectral theory to quantitative construction and analysis of quantum theoretic models is just a special case of relative entropic modelling. In consequence \cite{Kostecki:2011:AIP,Kostecki:2011:QIG}, it is possible to consider information geometry of spaces of non-commutative integrals as a new framework for kinematics of quantum information theory (in particular) and quantum theory (in general), replacing the traditional and semi-spectral frameworks based on Hilbert spaces and spectral theory \cite{vonNeumann:1932:grundlagen,deMuynck:2002} as well as the algebraic frameworks based on algebras of operators and spectral theory \cite{Segal:1947:postulates,Emch:1972,Haag:1992,Connes:1994}. 

\ \\{\small \textbf{Acknowledgements} I am indebted to P.~Gibilisco, F.~Hellmann, C.J.~Isham, A.~Jen\v{c}ov\'{a}, W.~Kami\'{n}ski, C.C.~Rodr\'{\i}guez, R.F.~Streater, and S.L.~Woronowicz for valuable discussions and remarks. I am especially grateful to professor S.L.~Woronowicz for the inspiring discussions on modular theory and non-commutative flow of weights. Many thanks to C.~Flori, J.~Vicary, and T.~Ko{\l}odziejski for their kind hospitality. Partial support by FNP \textit{Mistrz}, ESF \textit{QGQG} 1955, ESF \textit{QGQG} 2706, MSWiN \textit{182/N} QGG/2008/0, and NCN \textit{N N202} 343640 grants is acknowledged.}

{\scriptsize
\bibliographystyle{../rpkbib}

\bibliography{../rpkrefs}
}
\end{document}